\documentclass[twocolumn,prc,showpacs,preprintnumbers,amsmath,amssymb,superscriptaddress,nofootinbib,aps,10pt]{revtex4-1}

\usepackage{graphicx}
\usepackage{dcolumn}
\usepackage{bm}
\usepackage{booktabs}
\usepackage{placeins}

\begin{document}

\preprint{Submitted to Physical Review C Rapid Communications}

\title{Precision mass measurements of $^{125-127}$Cd isotopes and isomers approaching the $N=82$ closed shell}

\author{D. Lascar}
    \email[Corresponding author: ]{dlascar@triumf.ca}
    \affiliation{TRIUMF, 4004 Wesbrook Mall, Vancouver, British Columbia V6T 2A3, Canada}
\author{R. Klawitter}
    \affiliation{TRIUMF, 4004 Wesbrook Mall, Vancouver, British Columbia V6T 2A3, Canada}
    \affiliation{Max-Planck-Institut f\"{u}r Kernphysik, Heidelberg D-69117, Germany}
\author{C. Babcock}
    \affiliation{TRIUMF, 4004 Wesbrook Mall, Vancouver, British Columbia V6T 2A3, Canada}
\author{E. Leistenschneider}
    \affiliation{TRIUMF, 4004 Wesbrook Mall, Vancouver, British Columbia V6T 2A3, Canada}
    \affiliation{Department of Physics \& Astronomy, University of British Columbia, Vancouver, British Columbia V6T 1Z1, Canada}
\author{S.R. Stroberg}
    \affiliation{TRIUMF, 4004 Wesbrook Mall, Vancouver, British Columbia V6T 2A3, Canada}
\author{B.R. Barquest}
    \affiliation{TRIUMF, 4004 Wesbrook Mall, Vancouver, British Columbia V6T 2A3, Canada}
\author{A. Finlay}
	\affiliation{TRIUMF, 4004 Wesbrook Mall, Vancouver, British Columbia V6T 2A3, Canada}
    \affiliation{Department of Physics \& Astronomy, University of British Columbia, Vancouver, British Columbia V6T 1Z1, Canada}
\author{M. Foster}
    \affiliation{TRIUMF, 4004 Wesbrook Mall, Vancouver, British Columbia V6T 2A3, Canada}
    \affiliation{Department of Physics, University of Surrey, Guildford GU2 7XH, United Kingdom}
\author{A.T. Gallant}
	\email{Present address: Lawrence Livermore National Laboratory, Livermore, California 94550, USA}
    \affiliation{TRIUMF, 4004 Wesbrook Mall, Vancouver, British Columbia V6T 2A3, Canada}
\author{P. Hunt}
    \affiliation{Department of Physics, Colorado School of Mines, Golden, Colorado, 80401, USA}
\author{J. Kelly}
	\affiliation{Department of Physics, University of Notre Dame, Notre Dame, Indiana, 46556, USA}
\author{B. Kootte}
    \affiliation{TRIUMF, 4004 Wesbrook Mall, Vancouver, British Columbia V6T 2A3, Canada}
    \affiliation{Department of Physics \& Astronomy, University of Manitoba, Winnipeg, Manitoba R3T 2N2, Canada}
\author{Y. Lan}
    \affiliation{TRIUMF, 4004 Wesbrook Mall, Vancouver, British Columbia V6T 2A3, Canada}
    \affiliation{Department of Physics \& Astronomy, University of British Columbia, Vancouver, British Columbia V6T 1Z1, Canada}
\author{S.F. Paul}
    \affiliation{TRIUMF, 4004 Wesbrook Mall, Vancouver, British Columbia V6T 2A3, Canada}
    \affiliation{Ruprecht-Karls-Universit\"{a}t Heidelberg, D-69117 Heidelberg, Germany}
\author{M.L. Phan}
    \affiliation{TRIUMF, 4004 Wesbrook Mall, Vancouver, British Columbia V6T 2A3, Canada}
    \affiliation{Department of Physics \& Astronomy, University of British Columbia, Vancouver, British Columbia V6T 1Z1, Canada}
\author{M.P. Reiter}
    \affiliation{TRIUMF, 4004 Wesbrook Mall, Vancouver, British Columbia V6T 2A3, Canada}
    \affiliation{GSI Helmholtzzentrum f\"{u}r Schwerionenforschung GmbH, Planckstra{\ss}e 1, 64291 Darmstadt, Germany}
\author{B. Schultz}
    \affiliation{Department of Physics, University of Notre Dame, Notre Dame, Indiana, 46556, USA}
    \affiliation{TRIUMF, 4004 Wesbrook Mall, Vancouver, British Columbia V6T 2A3, Canada}
\author{D. Short}
    \affiliation{TRIUMF, 4004 Wesbrook Mall, Vancouver, British Columbia V6T 2A3, Canada}
    \affiliation{Department of Chemistry, Simon Fraser University, Burnaby, British Columbia V5A 1S6, Canada}
\author{J. Simonis}
	\affiliation{Institut f\"ur Kerphysik, Technische Universit\"at Darmstadt,
64289 Darmstadt, Germany}
	\affiliation{ExtreMe Matter Institute EMMI, GSI Helmholtzzentrum f\"ur
Schwerionenforschung GmbH, 64291 Darmstadt, Germany}
\author{C. Andreoiu}
    \affiliation{Department of Chemistry, Simon Fraser University, Burnaby, British Columbia V5A 1S6, Canada}
\author{M. Brodeur}
    \affiliation{Department of Physics, University of Notre Dame, Notre Dame, Indiana, 46556, USA}
\author{I. Dillmann}
	\affiliation{TRIUMF, 4004 Wesbrook Mall, Vancouver, British Columbia V6T 2A3, Canada}
    \affiliation{Department of Physics and Astronomy, University of Victoria, Victoria, British Columbia V8P 5C2, Canada}
\author{G. Gwinner}
    \affiliation{Department of Physics \& Astronomy, University of Manitoba, Winnipeg, Manitoba R3T 2N2, Canada}
\author{J.D. Holt}
    \affiliation{TRIUMF, 4004 Wesbrook Mall, Vancouver, British Columbia V6T 2A3, Canada}
\author{A.A. Kwiatkowski}
    \affiliation{Cyclotron Institute, Texas A\&M University, College Station, Texas 77843, USA}
    \affiliation{TRIUMF, 4004 Wesbrook Mall, Vancouver, British Columbia V6T 2A3, Canada}
\author{K.G. Leach}
    \affiliation{Department of Physics, Colorado School of Mines, Golden, Colorado, 80401, USA}
    \affiliation{TRIUMF, 4004 Wesbrook Mall, Vancouver, British Columbia V6T 2A3, Canada}
\author{J. Dilling}
    \affiliation{TRIUMF, 4004 Wesbrook Mall, Vancouver, British Columbia V6T 2A3, Canada}
    \affiliation{Department of Physics \& Astronomy, University of British Columbia, Vancouver, British Columbia V6T 1Z1, Canada}



\date{\today}


\begin{abstract}
We present the results of precision mass measurements of neutron-rich cadmium isotopes. These nuclei approach the $N=82$ closed neutron shell and are important to nuclear structure as they lie near doubly-magic $^{132}$Sn on the chart of nuclides. Of particular note is the clear identification of the ground state mass in $^{127}$Cd along with the isomeric state. We show that the ground state identified in a previous mass measurement which dominates the mass value in the Atomic Mass Evaluation is an isomeric state. In addition to $^{127/m}$Cd, we present other cadmium masses measured ($^{125/m}$Cd and $^{126}$Cd) in a recent TITAN experiment at TRIUMF. Finally, we compare our measurements to new \emph{ab initio} shell-model calculations and comment on the state of the field in the $N=82$ region.

\end{abstract}

\maketitle

\section{Introduction}
Long-lived isomeric states of short-lived nuclei are abundantly observed in the region of the Segre chart near the doubly magic nucleus $^{132}$Sn. This trend is largely due to low lying $\nu h_{11/2}$ and $\pi g_{9/2}$ configurations near several low-$j$ orbitals. In this region, the small number of valence particles or holes provides an ideal environment for performing detailed tests of modern shell-model calculations in heavy-mass ($A>100$) nuclei \cite{Walker1999}. To date, the nuclear physics community has expended great effort to study this region experimentally (e.g. \cite{Gorska2009,Yordanov2013,Hakala2012}), however a complete picture of these nuclei still requires more detailed and precise measurements. Perhaps one of the best systems to pursue such studies is that of the neutron-rich cadmium nuclei, as they are near the closed proton and neutron shells at $Z=50$ and $N=82$, respectively. The odd-$A$ isotopes of cadmium, near $N=82$ in particular, are all known to contain isomers with half-lives similar to that of their ground states \cite{Audi2012a}. However, in many cases the energies of these isomers and their spin assignments remain unknown \cite{ENSDF}. Further investigations are therefore required to better test current state-of-the-art shell-model calculations.

Previous investigations of these isomers have been primarily performed via $\gamma$-spectroscopy. Despite the high quality data provided, these techniques are only sensitive to differences between the initial and final states, and thus have some limitations. In fact, with one notable exception the energies and spin assignments of the ground and isomeric states of $^{125}$Cd and $^{127}$Cd are either unknown or assigned purely based on systematic trends in the Evaluated Nuclear Structure Data Files (ENSDF) \cite{ENSDF}. The lone exception is the energy of $^{125m}$Cd where the 2012 Atomic Mass Evaluation (AME2012) \cite{Audi2012,Wang2012} cites a single source \cite{Kankainen2013} that will be discussed at length later in this publication.

At this point we note that all mass comparisons in this publication are performed with respect to 2012 AME \cite{Audi2012,Wang2012} and not the 2016 iteration \cite{Huang2017,Huang2017a}. This is because the mass values in this publication were transmitted to the AME 2016 authors as a private communication (cited in that publication as 16La.A). It would be redundant to compare the masses in this publication to the values in an evaluation that already contain them.

The evolution of the $N=82$ shell gap, in particular, has been the subject of intense scrutiny over the years. It was initially predicted to be reduced or even quenched \cite{Dobaczewski1994,Dillmann2003}. Recent mass measurements of $^{129-131}$Cd show a $\sim 1$ MeV reduction in the shell gap energy \cite{Atanasov2015,Knobel2016}. Also, evidence of enhanced quadrupole collectivity was found in neutron-rich, even-$A$ Cd nuclei \cite{Caceres2009,Kautzsch2000}, which was extended to decay spectroscopy of odd-$A$ Cd (e.g. \cite{Naqvi2010}) and laser spectroscopy \cite{Yordanov2013}. The picture that emerged for $^{125}$Cd and $^{127}$Cd is one where there are two independent decay chains that are individually well understood but their relationship remains unconfirmed.

In addition to the existing experimental data, Penning Trap Mass Spectrometry (PTMS) provides an elegant and well established solution \cite{Blaum2012} to the problem of determining the energy separation between the ground state and the isomer. As long as a given state is sufficiently long-lived, its mass can be measured and the relative binding energy between the states can be inferred. With sufficiently high resolution (in this case, $\Delta m \leq 70$ keV/$c^2$) and a cocktail beam containing  the isomer and the ground state, both can be measured and their energy difference determined to high precision in the same experiment. This article reports PTMS measurements for $^{125g,m}$Cd, $^{126}$Cd, and $^{127g,m}$Cd.

\section{C\lowercase{d} Mass Measurements}
\label{sec:Cd}


\begin{table*}[ht]
  \centering
  \caption{Measured values of $^{125-127}$Cd, the half-lives, $t_{1/2}$, of the species in question and the ion production as measured by the ISAC operators. The production measurements do not differentiate between ground and isomeric states and so are listed as ground states. R is the ratio of the cyclotron frequency of $^{133}$Cs$^{13+}$ to the cyclotron frequency of the ion of interest. $\Delta$ is the mass excess in keV/c$^2$. $\Delta_{\mathrm{AME}}$ is (with the exception of the value for $^{125m}$Cd) the value from the 2012 AME \cite{Audi2012,Wang2012}. We do not compare to the 2016 AME \cite{Huang2017,Huang2017a} because the values in this publication were included as a private communication and cited as 16La.A. For $^{125m}$Cd the value is taken from the mass measurement made by JYFLTRAP \cite{Kankainen2013}. It should be noted that the ground state values in the AME for these nuclides are dominated by measurements reported in a publication by JYFLTRAP \cite{Hakala2012} indicating that they have outweighed all previous Q($\beta$)-derived mass values. ``Difference'' is the difference between the measured value at TITAN and the literature value. The two \textbf{boldface} values are of note because it appears that the ground state in the AME has been misidentified as either an isomeric state or an amalgam of the ground state and isomer dominated by the isomer. The value listed in the ``Difference'' column for $^{127m}$Cd is the difference between the measured TITAN value and the AME value for the ground state. The final column contains the mass value given in the FRDM(2012) \cite{Moller2016}. The symbol ``\#'' denotes that the value is extrapolated and unmeasured.}
    \begin{tabular*}{\textwidth}{@{\extracolsep{\fill}} cccccccc}
    \hline
    Species & $t_{1/2}$ \cite{Audi2012a} & Yields & R & $\Delta_{\mathrm{TITAN}}$ & $\Delta_{\mathrm{AME}}$ & Difference & $\Delta_{\mathrm{FRDM}}$ \\
          & (ms) & $s^{-1}$ &  & (keV/$c^2$) & (keV/$c^2$) & (keV/$c^2$) & (keV/$c^2$) \\
    \hline
    $^{125}$Cd$^{13+}$ & 680 & $1.4 \times 10^4$ & 0.93992254(22) & -73347(24) & -73348.1(2.9) & 1(24) & -73410 \\
    $^{125m}$Cd$^{13+}$ & 480 & - & 0.93992407(8) & -73157.9(9.0) & -73162(4) \cite{Kankainen2013} & 3.9(6.0) & - \\
    \hline
    $^{126}$Cd$^{13+}$ & 515 & $9 \times 10^3$ & 0.94745587(8) & -72260.7(7.0) & -72256.8(2.5) & -3.9(7.4) & -72460 \\
    \hline
    $^{127}$Cd$^{13+}$ & 370 & $4.75\times 10^3$ & 0.95500883(6) & -68743.4(5.6) & -68491(13) & \textbf{-253(14)} & -69010 \\
    $^{127m}$Cd$^{13+}$ & 200\# & - & 0.95501112(6) & -68460.1(4.7) & - & \textbf{30(14)} & - \\
    \hline
    $^{133}$Cs$^{14+}$ & Stable & - & 0.92856759(3) & -88072.1(3.0) & -88070.931(8) & 1.1(3.0) & -87740 \\
    \hline
    \end{tabular*}%
  \label{tab:values}%
\end{table*}%

We performed cadmium mass measurements at \mbox{TRIUMF's} Ion Traps for Atomic and Nuclear science (TITAN) facility \cite{Dilling2006}. Table \ref{tab:values} contains the compiled measured values from the TITAN Cd experiment along with recent literature values.  A $\sim 500$ MeV proton beam from the TRIUMF cyclotron was sent to the Isotope Separator and ACcelerator (ISAC) facility \cite{Dombsky2000} and impinged upon a UCx target. TRIUMF's Ion Guide Laser Ion Source (IG-LIS) \cite{Raeder2014} was used to produce Cd ions. IG-LIS electrostatically suppresses surface-ionized contaminants by a factor of $10^{5}-10^{6}$ and resonantly laser-ionizes elements of interest that diffuse out of the solid target in a neutral charge state. Ions from IG-LIS were sent through ISAC's magnetic mass separator which has a resolving power ($m/\Delta m$) of approximately 2,000 \cite{Bricault2014}. The continuous beam was cooled and bunched in TITAN's RadioFrequency Quadrupole cooler/buncher (RFQ) \cite{Brunner2012a} and singly charged ion bunches were sent to the TITAN Electron Beam Ion Trap (EBIT) \cite{Lapierre2010} for charge breeding. There TITAN takes advantage of a well-established program for increased precision in mass measurements of short-lived isotopes using highly charged ions \cite{Ettenauer2011}. $13^+$ and $14^+$ charge states of the ions were sent towards TITAN's Measurement PEnning Trap (MPET) \cite{Dilling2003} for a precision measurement and on their way contaminants selectively steered away by their times-of-flight in TITAN's Bradbury Nielsen Gate (BNG) \cite{Brunner2012}.

\begin{figure}[ht]
    \begin{center}
        \includegraphics[width=0.47\textwidth]{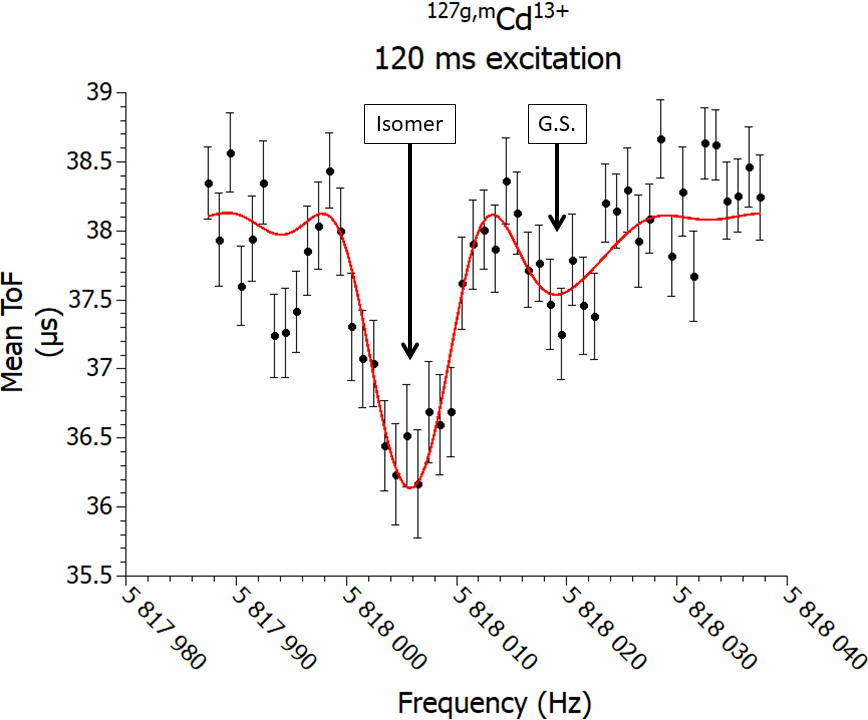}
        \caption{Typical ToF spectrum for $^{127}$Cd$^{13+}$. Both the ground state (shallower trough) and isomeric state (deeper trough) are easily visible (Color online).}
        \label{fig:spectra}
    \end{center}
\end{figure}

Inside MPET and its 3.7 T magnetic field, the ion bunch was further cleaned via the direct dipole excitation \cite{Blaum2007} of three likely isobaric contaminants, $^{A}$Cs$^{13+}$,  $^{A}$In$^{13+}$, and $^{A}$Cd$^{14+}$ where $A$ is the nucleon number of the cadmium isotope being investigated at the time. Subsequent scans of the dipole frequency made after those contaminant ions and the ion of interest were cleaned from MPET showed no discernible concentration of a single contaminant.

All TITAN mass measurements were made inside MPET performing the Time-of-Flight Ion Cyclotron Resonance (ToF-ICR) technique \cite{Bollen1990b}. The cyclotron frequency of an ion trapped in a strong magnetic field is
\begin{equation}
\nu_c = \frac{1}{2\pi}\frac{ze}{m}B
\end{equation}
where $m$ is the ion's mass, $z$ is the ion's charge state, $e$ is the elemental charge, and $B$ is the magnetic field strength.

The precision of ToF-ICR measurements is given by \cite{Bollen2001},
\begin{equation}
\label{eq:tof-icr-precision}
\frac{\delta m}{m} \propto \frac{m}{zBt_{\mathrm{rf}}\sqrt{N_{\mathrm{ion}}}},
\end{equation}
where $t_{\mathrm{rf}}$ is the amount of time the ion in question is excited via a quadrupole excitation in the Penning trap, and $N_{\mathrm{ion}}$ is the number of ions detected in a ToF spectrum (see Figure \ref{fig:spectra}) over the course of a measurement. $\delta m/m$ can be improved by measuring the mass of an ion in a higher charge state, $z$. TITAN makes use of an EBIT which generates such higher charge states via the process of electron impact ionization using a 100 mA electron beam with an approximate energy of 4 keV. A charge-state was selected that can be reached within a fraction of the lifetime of the ion of interest, and is as high as possible (see Eq. \ref{eq:tof-icr-precision}) while avoiding regions of high background in the time of flight spectrum generated by ionized residual gases. To maximize the amount of ions $N_z$ in the desired charge state, the charge breeding time was optimized such that $N_z/\sum N_i \gtrsim 0.25$, where $N_i$ is the number of ions in charge state $i\neq z$. For the cadmium measurements a charge state of $z=13$ was found to be optimal.

The quadrupole excitation times used for various measurement runs are listed in the bottom of Table \ref{table:constants}. They were chosen as a compromise between the higher resolution achieved with a long excitation time, and the higher ion losses over time due to radioactive decay and to interactions with residual gas in MPET. For $^{125}$Cd and $^{127}$Cd the ground and isomeric states were both observed in the trap simultaneously (see, for example, Figure \ref{fig:spectra}). Each measurement followed the same pattern. First, the cyclotron frequency of $^{133}$Cs$^{13+}$ was measured, followed by a measurement of the cyclotron frequency of the ion of interest. The pattern was followed so that there was always a $^{133}$Cs$^{13+}$ measurement before and after a measurement of an ion of interest to bracket the measurement and determine the uniformity of the magnetic field over time. This way, systematic shifts in the experimental system (primarily due to, but not limited to, a drifting magnetic field) could be monitored, accounted for, and minimized. 


\section{Data Analysis}
\label{sec:analysis}

Analysis of the data followed the same procedure outlined in other recent publications (e.g. \cite{Brodeur2009} for the basics of ToF-ICR measurements at TITAN including ion-ion interactions, \cite{Gallant2014} for basic treatment of ToF-ICR with HCI, and \cite{Klawitter2016} for statistical methods unique to HCI). In all cases, we only analyzed shots from MPET that contained 2 or fewer detected ions in order to minimize the effects of ion-ion interactions. Because of low statistics from the radioactive species, a count-class analysis \cite{Kellerbauer2003} was not performed on the full data set as there was not enough data to be split into even three classes. 

Because the precise value of the magnetic field in the trap can shift over time, we monitor this quantity by measuring the cyclotron frequency of a well-known calibrant ion, in this case $^{133}$Cs$^{13+}$ ($\delta m$ of $^{133}$Cs = 8 eV/$c^2$) \cite{Audi2012,Wang2012}. As long as the ions of interest and the calibrant ions probe the same region of the field, taking a ratio of the two frequencies will cancel the magnetic field's contribution to the calculation of the mass. Non-linear fluctuations in the magnetic field are minimized via the use of a pressure-regulating system and their contribution to the uncertainty of mass measurements has been found to be several orders of magnitude lower than the level of precision of these mass measurements. The atomic mass is then given by
\begin{equation}
M = \frac{z}{z_{\mathrm{ref}}}R \left(M_{\mathrm{ref}} - z_{\mathrm{ref}}m_e + BE_{\mathrm{ref}} \right) + zm_e - BE,
\end{equation}
where all subscripts ``ref'' denote properties of the reference ion ($^{133}$Cs$^{13+}$), $m_e$ is the mass of the electron, $BE$ is the total electron binding energy of the ion in question, and $M$ is used to separate the mass of the neutral atom from, $m$, the mass of the ion. The frequency ratio $R$ is given by,
\begin{equation}
R = \frac{\nu_{c,\mathrm{ref}}}{\nu_c}.
\end{equation}
For singly or doubly charged ions, the binding energy can be neglected for mass measurements made to keV/$c^2$-levels of precision, but for highly charged ions the electron binding energies exceed 1 keV and must be taken into account. The binding energies, along with other parameters used to make mass calculations can be found in Table \ref{table:constants}. To check our calibration during the experiment we measured $\nu_c$ of $^{133}$Cs$^{14+}$ and used that to determine the mass of $^{133}$Cs with the $^{133}$Cs$^{13+}$ ion as the calibrant.

\begin{table}[ht]
\centering
\caption{Constants used to make mass calculations. At the bottom of the table are the RF excitation times used in the various experimental runs.}
\begin{tabular}{lr}
    \hline
    \multicolumn{2}{c}{Binding energies \cite{Rodrigues2004}} \\
    \hline
    Cd$^{13+}$ & 1.539(10) keV \\
    Cs$^{13+}$ & 1.532(10) keV \\
    Cs$^{14+}$ & 1.849(10) keV \\
    \hline \\
    \hline
    Species & $t_{\mathrm{rf}}$ used (ms) \\
    \hline
    $^{125}$Cd & 400, 200 \\
    $^{126}$Cd & 100 \\
    $^{127}$Cd & 120, 170 \\
    \hline
\end{tabular}
\label{table:constants}
\end{table}

\begin{figure*}[ht]
    \begin{center}
        \includegraphics[width=\textwidth]{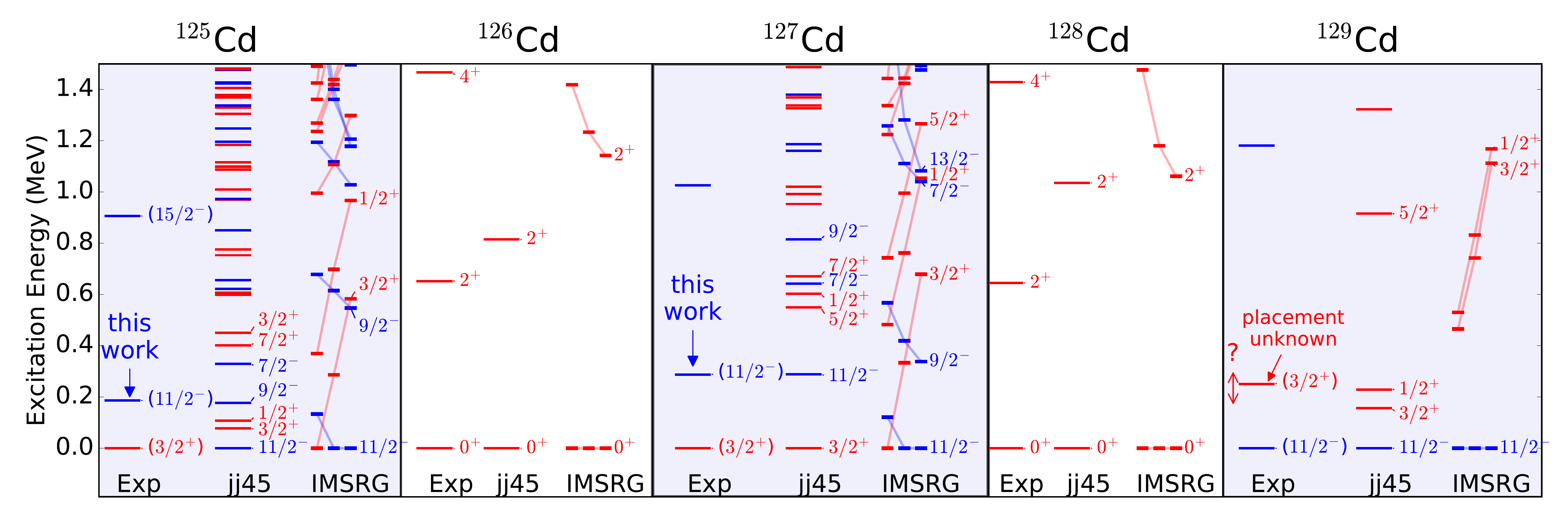}
        \caption{Experimental level schemes for $^{125-129}$Cd compared with phenomenological shell-model predictions using the jj45 interaction \cite{Dillmann2003} and {\it ab initio} results obtained with the valence-space IMSRG \cite{Tsukiyama2012,Bogner2014,Stroberg2016,Stroberg2017}. Experimental values from outside this work were taken from the National Nuclear Data Center \cite{NNDC2016}. Positive parity states are indicated with red lines while negative parity states are indicated with blue lines. Black lines indicate unknown parity. The energy of the $3/2^+$ state in $^{129}$Cd relative to the $11/2^-$ state is unknown experimentally and so its placement in the level scheme is arbitrary. The IMSRG results display a series of $e_{\mathrm{max}}/E_{3\mathrm{max}}$ truncations (from left to right) 14/14, 14/16, 14/18 (see text for details). No indication of convergence is observed (Color online).}
        \label{fig:level}
    \end{center}
\end{figure*}


Previous work in this region was performed at JYFLTRAP in Jyv\"{a}skyl\"{a} ($^{125g,m}$Cd, $^{126}$Cd, and $^{127g,m}$Cd) \cite{Hakala2012,Kankainen2013} and ISOLTRAP at CERN ($^{126}$Cd) \cite{Breitenfeldt2010b}. The mass values given in the 2012 AME \cite{Audi2012,Wang2012} reflect the dominance of Penning trap mass measurements. All Q($\beta$)-derived mass values in this region that were used in the 2003 AME \cite{Audi2003b,Wapstra2003b} have been outweighed in the 2012 iteration. In this light, measured values of cadmium isotopes are of considerable interest as a check and confirmation of previous work. While the TITAN measurements have good agreement with the 2012 JYFLTRAP measurements (less than $1\sigma$ for $^{125}$Cd, $^{125m}$Cd, $^{126}$Cd, and the important calibrant check of $^{133}$Cs$^{14+}$) there is a sizable disagreement between TITAN's and JYFLTRAP's measurements of $^{127}$Cd. The ground state that TITAN measured differs from the 2012 value by 253 keV/c$^2$ ($18\sigma$) while the isomer that TITAN measured differs by only 30 keV/$c^2$ ($2.1\sigma$) from the AME/JYFLTRAP ground state value. 

It seems possible that the ground state identified and measured by JYFLTRAP \cite{Hakala2012} was the isomeric state. A later publication on isomeric states in this mass region by the same group notes that ``the higher-spin states dominate the resonances of Cd and In isotopes'' \cite{Kankainen2013}. We therefore report isomer energies for $^{125}$Cd and $^{127}$Cd of 190(26) keV and 283.3(5.6) keV, respectively. Both are thought to be $11/2^{-}$ states (based on systematic arguments) and for $^{127m}$Cd this is the first measurement of the excitation energy. For $^{125}$Cd, our value for the energy difference between the ground state and the isomer of 190(26) keV agrees with the literature value of 185(5) keV \cite{Kankainen2013}.

With respect to the two ISOLTRAP mass measurements of $^{126}$Cd that were published in 2010 \cite{Breitenfeldt2010b}, TITAN's measured mass excess of -72,260.7(7.0) keV/$c^2$ is in agreement with both the -72,256.5(4.2) keV/c$^2$ value (from ``Experiment 1'') and the -72,266(14) keV/c$^2$ value (from ``Experiment 3'').


Attempts to measure the masses of more neutron-rich Cd nuclides were unsuccessful due to both low production of the isotope(s) of interest and high production of contaminants that could not be removed using standard dipole cleaning inside the precision Penning trap. Future TITAN mass measurement experiments will include a new Multi-Reflection Time-of-Flight mass spectrometer (MR-ToF) \cite{Jesch2015} that will provide sufficient mass-resolving power to allow TITAN to explore more masses in this region. The first steps will be to confirm the work of the FRS-ESR facility at GSI \cite{Knobel2016} and the more precise ISOLTRAP measurements \cite{Atanasov2015} at and just beyond the N=82 closed neutron shell. Afterwards, the TITAN MR-ToF coupled to TRIUMF's new ARIEL facility \cite{Dilling2014} will grant TITAN access to still more neutron-rich nuclides to study.

\section{Discussion}
\label{sec:conclusion}

In order to determine the size of the systematics of the structure of cadmium isotopes in this region, we have performed theoretical calculations of the low-lying spectra of $^{125-129}$Cd and compared the results with the existing experimental data in Figure~\ref{fig:level}.
We first performed a standard shell-model calculation with the NuShellX shell-model code~\cite{Brown2003}, using the jj45 effective interaction~\cite{Dillmann2003} in a valence space consisting of the ${(0g_{7/2},1d_{5/2},1d_{3/2},2s_{1/2})}$ orbits for protons, and ${(0h_{11/2},1f_{7/2},1f_{5/2},2p_{3/2},2p_{1/2})}$ for neutrons, above a $^{78}$Ni core.
While we have noted that the tentative spin assignments for the experimental levels are essentially all made based on systematics and comparison with shell-model calculations, we must simultaneously point out that the separation of ground state and isomer is small for the odd-mass isotopes, and the calculated ground state spins alternate between $11/2^{-}$ and $3/2^+$ and back again when going from $^{125}$Cd to $^{127}$Cd to $^{129}$Cd.

Systematics here do not justify a tentative spin assignment, and therefore the spins must be measured directly. Yordanov \emph{et al.} recently performed isomeric frequency shift measurements in a collinear laser spectroscopy experiment \cite{Yordanov2013,Yordanov2016}. While an energy difference of $\sim 250$ keV would produce a frequency difference of $\pm 1.3$ MHz \cite{Yordanov2016a}, that is approximately half the error stated in their publication \cite{Yordanov2016} so laser spectroscopy alone cannot illuminate the spin assignments either.



We also performed exploratory {\it ab initio} calculations using the valence space In-Medium Similarity Renormalization Group (IMSRG) approach~\cite{Tsukiyama2012,Bogner2014,Stroberg2016,Stroberg2017}.
Recently, one of the chiral interactions developed in Refs.~\cite{Hebeler2011,Simonis2016}\footnote{There labeled EM1.8/2.0} has been found to reproduce binding energies and spectroscopy for a wide range of nuclei, from the $p$-shell through the $fp$-shell~\cite{Simonis2017}.
We work in a harmonic oscillator single-particle basis with frequency $\hbar\omega=16$ MeV, truncated by $2n+\ell \leq e_{\mathrm{max}}$, making an additional cut on the three-body matrix elements $e_1+e_2+e_3\leq E_{3\mathrm{max}}$.
We then decouple the same valence space as was used in the standard shell-model calculations described above.
When using chiral interactions softened by the similarity renormalization group (SRG)~\cite{Bogner2007,Roth2014b}, a truncation $e_{\mathrm{max}}=14$, $E_{3\mathrm{max}}=14$ is typically sufficient to converge medium-mass nuclei lighter than $^{40}$Ca.  For this particular interaction, $E_{3\mathrm{max}}=16$ is sufficient up to $^{56}$Ni, but by $^{78}$Ni results are not fully converged with $E_{3\mathrm{max}}=18$, the current computational limit \cite{Simonis2017}. We would therefore not expect calculations of nuclei in the region near $^{132}$Sn to be converged. This is what was found  Ref.~\cite{Binder2014}, and this has been confirmed for the present interaction.

Regardless, there was a reasonable expectation that while absolute binding energies were not converged, excitation energies, being a relative quantity, would converge more rapidly. As seen in Figure~\ref{fig:level}, however, this is not the case.
The columns labeled ``IMSRG'' show the results of a series of truncations in $e_{\mathrm{max}}=14$ and $E_{3\mathrm{max}}=14,16,18$.
Note that in order to include an interaction between three $h_{11/2}$ neutrons ($\ell=5$), one must have at least $E_{3\mathrm{max}}=15$. 
While levels of a given parity appear reasonably converged, there is no sign of convergence between negative and positive parity states.

\begin{figure}[htb]
    \begin{center}
        \includegraphics[width=0.47\textwidth]{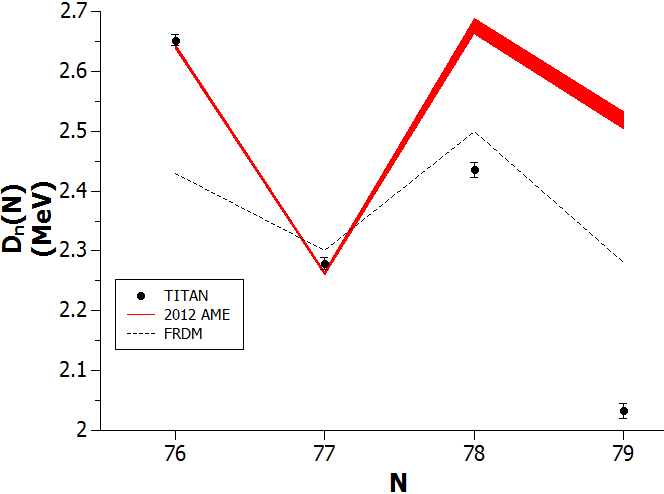}
        \includegraphics[width=0.47\textwidth]{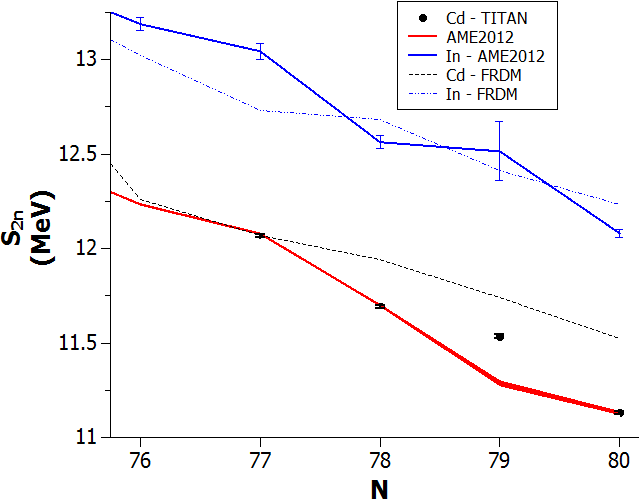}
        \caption{\emph{Top:} Shifts in the pairing gap relative to the values in the 2012 AME \cite{Audi2012,Wang2012} and the 2012 update to the FRDM \cite{Moller2016}. \emph{Bottom:} $S_{2n}$ calculated from recent TITAN mass measurements (black points) plotted relative to AME 2012 values for Cd (red band) and In (blue band). In both cases, only ground state masses were used (Color online).}
        \label{fig:structure}
    \end{center}
\end{figure}

Taking the ground state masses measured in this work, it is possible to observe shifts in the pairing gap energies, given in \cite{Brown2013} as
\begin{equation}
D_{n}(N) = (-1)^{N+1} \left[S_{n}(Z,N+1) - S_{n}(Z,N)\right].
\end{equation}
The topmost graph of Figure \ref{fig:structure} shows the results of those calculations with the new TITAN masses (combined with 2012 AME masses for the low-$A$ Cd values that were unexplored in this work), the 2012 AME masses, and masses from the 2012 update to the Finite Range Droplet Model (FRDM2012) \cite{Moller2016}. Included in Table \ref{tab:values} are values from the FRDM(2012), which is germane since the shell-model results cannot provide robust mass predictions in this region.

$D_n(N)$ depends on the pairing correlation intensity. Shifts of 200-500 keV are observed between the AME results, which take the isomeric state of $^{127}$Cd as the ground state, and those calculated from the new TITAN mass measurements. As a result we observe a relative suppression of the pairing strength in the $^{127}$Cd ground state, in contrast to the prediction of the FRDM mass model. When we also consider the uncertainty in ground-state spin assignments from the shell-model calculations, the neutron-rich cadmium isotopes emerge as a new challenge for existing theory in this region.


The lower panel in Figure \ref{fig:structure} shows two-neutron separation energies ($S_{2n}$) with the same set of mass values as in the top panel. We note that with the new mass value for $^{127}$Cd, the trend of cadmium $S_{2n}$ values is now similar to that in the adjacent indium isotopes, obtained from the 2012 AME. Any hints of changes in the nuclear structure that were suggested by earlier $S_{2n}$ analyses around $N=79$ appear to be suppressed, but the remaining uncertainty in $S_{2n}$ of $^{128}$In prevents a stronger conclusion at this time.

\section{Conclusion}

The measurement of $^{125-127}$Cd ground state and $^{125m,127m}$Cd isomeric state masses point to the importance of confirmation measurements in scientific endeavors. The authors of the 2012 AME were justified in outweighing the Q($\beta$) measurements used in determining the mass of $^{127}$Cd in the 2003 AME in favor of the newer 2012 measurement. The value had been basically unchanged since the 1993 iteration \cite{Audi1993}, there was no indication of any discrepancy in the 2012 measurement, and the precision cited was more than 5 times smaller. Our measurement of both the ground state and the isomer of $^{127}$Cd show that the mass measurement community was, in this case, mistakenly measuring and refining the precision of the isomeric mass.

The shell-model and \emph{ab initio} IMSRG calculations presented here casts doubt on the spin assignments in neutron-rich cadmium isotopes. Systematic arguments do not appear to be sufficient as the pairing energies in question could be enough to bring $3/2^+$ states higher than $11/2^-$ states. More experimental data in this region are necessary to definitively assign spin values to states and bring clarity to the nuclear structure questions that this work has posed. Concurrently, further development is required on the theoretical side, as jj45 is one of the few available phenomenological shell-model interactions in this region of the nuclear chart.  In addition we have seen that state-of-the-art \emph{ab initio} calculations are hindered by current computational limits and will face significant  challenges to reach convergence in this heavy-mass region.


\section*{Acknowledgement}
TRIUMF receives federal funding via a contribution agreement with the National Research Council of Canada (NRC). This work was partially supported by the Natural Sciences and Engineering Research Council of Canada (NSERC), the Canada Foundation for Innovation (CFI), the US National Science Foundation under Grant PHY-1419765, the Deutsche Forschungsgemeinschaft (DFG) under Grant FR 601/3-1, Brazil's Conselho Nacional de Desenvolvimento Cient\'{\i}fico e Technol\'{o}gico (CNPq), and the BMBF under Contract No. 05P15RDFN1.

The valence-space IM-SRG code used in this work makes use of the Armadillo \texttt{C++} library \cite{Armadillo}. Computations were performed with an allocation of computing resources at the J\"ulich Supercomputing Center.
DL wishes to thank C. Jannace and F. Friend for editing support.

\bibliography{library}%
\end{document}